# Internal effective field sources for spin torque nano pillar oscillators

Gino Hrkac[1], Thomas Schrefl[1], Julian Dean[1], Alexander Goncharov[1], Simon Bance[1],

Dieter Suess[2], Josef Fidler[2]

[1]Department of Engineering materials, University of Sheffield

Sir Robert Hadfieldbuilding, Mapping Street, S13JD Sheffield, UK

[2]Institue of Solid State Physics, Vienna University of Technology

Wiedner Haupt Strasse 8-10, 1040 Vienna, Austria

Abstract

In this paper we numerically conduct micromagnetic modelling with an expended micromagnetic model that includes the spin torque term and an impedance model to investigate methods to replace external field sources with internal ones and to investigate its tuneability on nanopillar geometries. We present results for three methods: interlayer coupling, large perpendicular anisotropy and magnetostatic coupling. The internal field sources are evaluated as function of frequency shift with current, its dependency on temperature and are tested against analytical predictions.

Introduction

Spin-torque high frequency microwave nanocontact-oscillators of ferromagnetic thin films have attracted interest, as monochromatic onboard gigahertz-frequency microwave sources for integrated electronic circuits. It has been theoretically predicted [1-3] and experimentally observed [4-6] that spin-polarized currents in magnetic nanostructure devices may excite steady state microwave magnetization precessions in ferromagnetic thin films. In these experiments an external field is applied to create a suitable resonance condition and to increase the tuneability of the frequency as function of current.

One of the main requirements of spin-transfer oscillations at high frequencies (e.g. 10 GHz and beyond) is the necessity of a large magnetic field. From a device point-of-view, the requirement of large magnetic fields imposes many constraints that are not compatible with CMOS integrated nano oscillators, i.e. that it is nanometric in size. A mean to avoid large external fields is to use large internal effective fields that are generated through coupling the magnetic free layer to other magnetic layers.

Since the first ferromagnetic resonance (FMR) experiments of W. Stoecklein et al. [7] in 1988 on exchange-biased NiFe/IrMn bilayers, there has been a large body of experimental and theoretical work that show a shift in resonance frequencies of the ferromagnetic layer. A spectacular example comes from an experiment of J. McCord et al. [8], in which they investigated the role of the antiferromagnetic film thickness on the ferromagnet resonance frequency. For a fixed ferromagnetic NiFe film thickness of 40 nm, they varied the antiferromagnetic FeMn film thickness between 0.2 nm and 4.3 nm and found that the zero-field FMR frequency shifted from 0.8 GHz to 1.9 GHz correspondingly. One can extrapolate these results to the thinner films envisaged in this paper which would be of the order of 4-5 nm in thickness. As such, in light of the interfacial nature of exchange bias, we expect that a ten-fold decrease in the ferromagnetic film thickness would lead to a ten-fold increase in the FMR frequency for the same materials, i.e. 8-19 GHz. Kuanr et al. have shown that such effects are also possible in NiFe (F)/NiO (AF) multilayers [9]. Another means of generating effective internal fields is to couple the active ferromagnet to another ferromagnet through an intermediate spacer layer. One advantage of this scheme over exchange bias is that the effective interlayer coupling can be quite large. For example, in Co/Ru/Co structures [10], the effective coupling field can attain a few Tesla. Since the tuneability of ferromagnetic precession frequencies are typically in the order of 28 GHz/Tesla, one can immediately see the benefit of having Tesla fields to generate precessional frequencies in the tens of GHz. A further approach to include internal effective fields is to use a material with a large perpendicular anisotropy. In this context, perpendicular refers to the direction normal to the film plane. In a recent calculation of Lee et al. [11], it is demonstrated that steady magnetization precession with frequencies between 1 and 20 GHz are possible in the absence of external magnetic fields. Furthermore, spin torque switching and current induced oscillations were recently demonstrated experimentally for a perpendicular geometry [12], where a Co/Ni multilayer with perpendicular anisotropy was used as the free layer.

In this paper we perform a numerical study with an expended micromagnetic model that includes the spin torque term and an impedance model to investigate methods to replace the external field sources with internal ones and to investigate its tuneability on nanopillar geometries. We study three main possibilities: interlayer coupling, large perpendicular anisotropy and magnetostatic coupling.

Theory

The solution of an electromagnetic problem including nonlinear magnetization dynamics consists of solving the static Maxwell equations and the Landau-Lifshitz-Gilbert equation modified with the Slonczewski term. The LLGS equation is solved with appropriate initial and boundary conditions. We use a micromagnetic approach starting from the total magnetic Gibbs free energy $E_{tot}$

$$E_{tot} = \int_V \left( \frac{A}{M_s^2} (\nabla \mathbf{M})^2 + E_{anis} - \mu_0 \mathbf{M} \left( \mathbf{H}_a + \frac{\mathbf{H}_d}{2} \right) \right) dV \qquad (1)$$

which is discretized using a finite element method that is solved using a hybrid finite element/boundary element method [14, 15]. The first term in equation 1 is the exchange energy followed by the anisotropy energy density, the applied field $\mathbf{H}_a$ and the magnetostatic energy with $\mathbf{H}_d$ being the demagnetizing field, with $\mu_0$ being the permeability of free space. The nonlinear magnetization dynamics including spin torque and thermal effects is described by a modified stochastic Landau-Lifshitz-Gilbert equation, which is written in the following dimensionless form

$$(1+\alpha^2) \frac{d\mathbf{m}}{d\tau} = -\mathbf{m} \times (\mathbf{h}_{eff} + \mathbf{h}_{th}) - \alpha \mathbf{m} \times (\mathbf{m} \times (\mathbf{h}_{eff} + \mathbf{h}_{th})) - \mathbf{N} \qquad (2)$$

with $\mathbf{N}$ being the spin transfer torque describing the effects caused by the spin polarized current and $\mathbf{h}_{th}$ is a random thermal fluctuation field. The magnetization vector $\mathbf{m} = \mathbf{M}/M_s$ is the magnetization $\mathbf{M}$ normalized by the saturation magnetization $M_s$, and is assumed to be spatially nonuniform, α is the Gilbert damping constant, τ the time measured in units of $(|\gamma| M_s)^{-1}$ (γ is the gyromagnetic ratio) and $\mathbf{h}_{eff} = \mathbf{H}_{eff}/M_s$ is the normalized effective field. The effective field is the negative variational derivative of the total magnetic Gibbs free energy density. The spin transfer torque **N** in equation 2 is given in the asymmetric form [16] and can be written as

$$\mathbf{N} = \frac{\hbar J_e}{2ed} \frac{1}{\mu_0 M_s^2} g_T \mathbf{m} \times (\mathbf{m} \times \mathbf{p}) = \frac{1}{2} \frac{J_e}{J_p} g_T \mathbf{m} \times (\mathbf{m} \times \mathbf{m}_p) \qquad (3)$$

Here $J_e$ is the electron current density, $J_p = (\mu_0 \, e \, d \, M_s^2 / \hbar)$ the characteristic current density of the system ($\hbar$ is the Planck constant, $e$ the electron charge, $d$ the free layer thickness) and $\mathbf{m}_p$ is the unit vector in the direction of the magnetization in the reference layer. The scalar function $g_T$ describes the angular dependency of the spin torque term and is commonly defined as follows [17]

$$g_T(\theta) = \left( \frac{8 P^{3/2}}{3(1+P)^3 - 16 P^{3/2} + (1+P)^3 \cos(\theta)} \right) \qquad (4)$$

Where P is the spin polarization factor and $\cos(\theta) = \mathbf{m} \cdot \mathbf{m}_P$ is the projection of the magnetization on the free layer **m** and the fixed layer $\mathbf{m}_P$. The spin current polarization is expressed by the constant P with values ranging from $0 \leq P \leq 1$. In the following simulation we assumed a value of P = 0.2.

In order to take thermal activation into account, equation 2 was expanded with a stochastic thermal field which is added to the effective field. This field accounts for interaction effects of the magnetization with macroscopic effects like phonons, conducting electrons, etc, which cause fluctuation in the magnetization dynamics [18]. As the contributing effects posse a large number of degrees of freedom, the thermal field is described as a Gaussian random distribution with the following statistical properties

$$\langle \mathbf{h}_{th,i}(t) \rangle = 0 \qquad (5)$$

Meaning that the average of the thermal activation taken over all realizations vanishes in each direction of space $i \in \{x, y, z\}$. The variance is given by

$$\langle \mathbf{H}_{th,i}(t); \mathbf{H}_{th,j}(t') \rangle = 2D\delta_{ij}\delta(t-t') \qquad (6)$$

Equation 6 is comparable to the fluctuation-dissipation theorem. It relates the strength of the thermal fluctuations to the dissipation due to damping. The assumption that the different components of the system are uncorrelated is expressed by the Kronecker δ whereas the Dirac δ expresses that the autocorrelation time of the thermal field is much shorter than the response time of the system. The strength of the thermal fluctuation D is derived from the fluctuation-dissipation theorem [19]

$$D = \frac{\alpha k_B T}{\gamma \mu_0 M_s V_i} \qquad (7)$$

With α being the damping constant, $k_B$ the Boltzmann constant, T the temperature, γ the gyromagnetic ratio, $M_s$ the saturation magnetization and $V_i$ the volume surrounding the node i of the finite element mesh.

The stochastic LLGS equation is solved by separating the deterministic and the stochastic contribution. By rearranging equation 2 on derives the following expression

$$(1+\alpha^2)\frac{d\mathbf{m}}{d\tau} = -\mathbf{m} \times \mathbf{h}_{eff} - \alpha \mathbf{m} \times (\mathbf{m} \times \mathbf{h}_{eff}) - \mathbf{N}$$
$$-\mathbf{m} \times \mathbf{h}_{th} - \alpha \mathbf{m} \times (\mathbf{m} \times \mathbf{h}_{th}) \qquad (8)$$

which shows that it is a Langevin type stochastic differential equation with multiplicative noise.

The nonlinear differential equation 8, is solved with a semi-implicit method where the right hand side of equation 8 is evaluated with a mid-point rule in the time domain and the final set of non linear equations are solved by functional iteration [20].

Interlayer coupling

One possibility to generate an effective internal field is to couple the active frerromagnetic layer to another ferromagnet through an intermediate spacer layer. We perform numerical simulations to demonstrate that the inclusion of additional internal fields does indeed lead to an increase in the oscillation frequencies. Here we simulate an elliptical shaped trilayer stack (Co 40 to 30 nm/Ru 0.8 nm/ Co 3nm) which consists of typical synthetic antiferromagnetic components (Co and Ru). Due to the 1/d term in the spin torque N, the spin current induced motion in the thicker layer is negligible. The thicker (40 to 30 nm) Co layer serves as the spin polarizer while the thinner (3nm) Co layer serves as the free layer and the Ru layer induces a strong interlayer coupling. The value of the magnitude of the interlayer coupling, and thereafter the effective field, can be tuned by varying the Ru layer thickness [13]. The interlayer coupling is an interface effect meaning that the internal effective field is inversely proportional to the ferromagnet film thickness. In addition, the overall magnitude of the interlayer coupling is also dependent on the thickness of the spacer layer. The combination of these two properties allows one to tailor the internal effective fields in a controlled fashion by adjusting the thickness of the ferromagnet and the spacer layer.

Based on the above consideration a micromagnetic model of Co/Ru/Co nanopillar was used to calculate V(t ,I, T, geometry). First the aspect ratio and layer geometry was optimized. The stray field may cause the formation of a vortex in the thicker Co layer. In a first attempt to avoid the vortex formation, we introduced an elliptic nano-pillar with an aspect ratio of 1.44 and a width of 100 nm. The 30 nm thick Co layer was pinned with an antiferromagnet leading to an average bias field of 11.7 mT. Shape anisotropy and bias field are sufficiently high to suppress vortex formation for zero current. However, an Oersted field produced by a current of 36 mA caused the formation of a vortex. A mean to avoid the vortex formation is to increase the exchange bias field. But even a reduction of the Co layer thickness to 20 nm which increased the exchange bias field to 17.5 mT could not suppress the vortex formation. Only after changing the aspect ratio to 2.25 a single domain state that remained stable in high currents was found.

Once a stable single domain state was found, we performed simulations to investigate the voltage output as function of temperature and the oscillation frequency as function of current and Temperature to account for Joule heating. The total simulation time was 200 nanoseconds and the temperature was T = 0 K and T = 450 K. For the simulation a spin polarization of 0.2 and a GMR ration of 2% were assumed [13].

First, the voltage was calculated as a function of time for one current, 36 mA, see figure 1, to quantify the voltage output range. It can be seen that the voltage output (quantitative values) is within the range of CMOS compatible values. Further it is shown that Joule heating (comparing the results of 0 and 450 K simulations) only imposes a small change in the output voltage amplitude.

To quantify the oscillation frequency shift with current we performed simulations with varying current at T=0 K, starting at 32 mA and increasing it up to 42 mA in 1 mA steps, see figure 2. By analyzing the oscillation signal and plotting the frequency as function of current and power, two things can be observed: first a red shift of oscillation frequency occurs with increasing current; second the power increases with current. But once it reaches higher currents, 36 mA and above, a clear broadening of the line width of the peaks can be observed which indicates that the oscillation behaviour is more non-uniform, see figure 2. In order to account for Joule heating the simulation were repeated with a temperature of 450 K. As shown in Figure 3, the thermal effects lead to a small shift of the frequencies.

Perpendicular Anisotropy

The second proposed method to include internal effective fields is to use a material with a large perpendicular anisotropy. In this context, perpendicular refers to the direction normal to the film plane. The anisotropy involved is necessarily very large, because it must first overcome the large demagnetizing field of the ferromagnet to provide a stable magnetic configuration at a finite angle out of plane. As such, the effective fields generated by a perpendicular anisotropy are of the order of a few Tesla, which translate to resonance frequencies in the tens of GHz.

The use of a perpendicular material can also entail different spin-transfer geometries. For example, one can envisage a polarizing layer with magnetization in the film plane and an active layer with a magnetization out of plane. The interplay between the perpendicular anisotropy and the demagnetizing field lead to a magnetization tilted slightly out of plane. One immediate advantage of this geometry is that there is automatically a large angle between the active layer and the polarizing layer, which facilitates a large torque due to spin-transfer without the need of initial fluctuations as in the case of collinear transfer.

Aside from the benefits of large anisotropy fields for the precession frequencies, this configuration also permit spin-transfer effects to occur at lower currents and zero applied fields. In a recent calculation of Lee et al. [11], it is demonstrated that steady magnetization precession with frequencies between 1 and 20 GHz are possible in the absence of external magnetic fields.

Here we demonstrate stable oscillations numerically for the following layer geometry: The polarizer is a 20 nm thick CoFe layer, the spacer layer is a 3 nm thick Cu layer, and the free layer is a 6 nm thick Co/Ni multilayer. The perpendicular anisotropy of Co/Ni multilayers can be tuned in the range from 30 to 400 kJ/m$^3$ [21]. For the following calculations we assumed a perpendicular anisotropy in the free layer of Ku = 200 kJ/m$^3$.

First we investigate the frequency change of the oscillator as function of applied current, which gives the tuneability range of the oscillator. The applied current starts at 2 mA and increases to 5 mA in 0.5 mA steps. We calculate the change of magnetization in time, and by making a Fourier analysis of the signal we calculate the power spectrum density as function of frequency and current, see figure 4. As the current is increased from 2 mA to 5 mA the frequency shifts to lower frequencies. To account for Joule heating, we performed simulations with 150, 300 and 450 K, see figure 5.

From the calculations two effects can be observed as the temperature is increased: A red shift in frequency and the oscillation modes become more non-uniform due to an increased noise background which is reflected in the broadening of the frequency peaks. This perfectly agrees with spin wave theory published by Kim [22]. Further, the line width, Δf, decreases with increasing spectral line intensity, $S_0$. Figure 6 give the line width as a function of the current and as a function of $S_0$. To compare this sample with the interlayer exchange coupled we calculate the voltage output for the oscillator stack for an applied current of 4mA and 3 different temperatures 150, 300 and 450 K, see figure 7. It can be seen from figure 7 that the temperature increase results in a blue shift in frequency but does not influence the strength of the signal (small change in amplitude).

Magnetostatic interaction field

The third possibility to introduce an effective internal field is with a magnetostatic interaction field. The magnetostatic field created by the reference layer may be used as a field to tune the oscillations of the free layer. In this case we also introduce an alternative way to stabilize the reference layer. As opposed to shape anisotropy we use a synthetic antiferromagnet. If the two layers of the synthetic antiferromagnet have different thicknesses or different magnetization the synthetic antiferromagnet will create a magnetic field. We used a IrMn (7nm)/ CoFe (1.8nm)/ Ru (0.8nm)/ CoFe (3.8nm) structure that will create a field of about 20 mT in the free layer. The distance between the synthetic antiferromagnet and the free layer is 3 nm. The circular nano-pillar has a diameter of 120 nm, meaning, that there is no shape anisotropy that stabilizes the magnetization. As a consequence thermal noise should be more severe and thermal fluctuations will drastically broaden the line width. To characterize this system, we investigate the frequency and the line width as function of current and temperature. The applied current is increased from 9 mA to 13 mA and the results are shown in figure 8.

In the left graph in figure 8, a red shift in frequency as function of current can be observed. The frequency changes from 7.4 GHz to 5.3 GHz by varying the current from 9 to 13 mA, which results in a tuneability of 0.525 GHz/mA. The simulations were performed with 150 and 300 K to investigate the effect of Joule heating on the frequency change and line width. From figure 8, it can be seen that a rise in temperature from 150 to 300 K has a small effect on the frequency change of the system but becomes more dominant for the line width. To investigate the line width as function of temperature in more detail, we performed simulations with 300 and 450 K at 12 and 13 mA, see figure 9.

Conclusions

We have presented three scenarios for the generation of internal effective fields for use in nano-pillar systems. The role of the internal field is twofold. First, the internal field allows for the possibility of high-frequency oscillations. Second, it negates the need for a large external field, the generation of which would neither be feasible nor desired on submicronic dimensions.

From the survey of existing experimental results and theoretical simulations of the three potential internal field sources, we find sufficient compelling evidence to support the use of internal effective fields such that the systems may operate at zero or low applied fields.

The micromagnetic simulations clearly indicate the high aspect ratio elliptical shaped nanopillars have to be used in the absence of an applied field. Otherwise the Oersted field from the current will lead to the formation of a vortex. Vortices may occur either in the reference layer or in the free layer. If the aspect ratio is too low, vortices may form immediately after a current is applied. For intermediate aspect ratios, vortices may be induced by thermal fluctuations or spin wave instabilities only after several nanoseconds of stable operation of the nanopillar. For currents up to 5 mA (current densities of 4.42 x $10^{11}$ A/m$^2$) an aspect ratio of 1.44 was sufficient. An aspect ratio of 2.25 was required, for currents in the range of 30 mA to 42 mA (current densities in range of 2.65 x $10^{12}$ A/m$^2$ to 3.71 x $10^{12}$ A/m$^2$).

Of the three effective fields considered, exchange biasing and interlayer coupling appear to be simpler to implement, as these already feature in some way in modern magnetoresistive stacks and do not require non-standard materials. The use of large perpendicular anisotropies, however, leads to a different geometry for spin-transfer that may prove to be advantageous for the zero- or low-field constraint.

This work was supported by the European Communities programs IST STREP, under Contract No. IST-016939 TUNAMOS.

References

1 J.C. Slonczewski, J. Magn. Magn. Mater. **159**, L1 (1996).

2 L. Berger, Phys. Rev. B **54**, 9353 (1996).

3 J.C. Sloncewski, J. Magn. Magn. Mater. **195**, L261 (1999).

4 S.I. Kiselev, J.C. Sankey, I.N. Krivorotov, N.C. Emley, R:A: Buhrman, and D.C: Ralph, Nature (London) **425**, 380 (2003).

5 W.H. Rippard, M.R. Putfall, S. Kaka, T.J. Silva, And S.E. Russek, Phys. Rev. B **70**, 100406(R) (2004).

6 W.H. Rippard, M.R. Putfall, S. Kaka, S:E. Russek, and T.J. Silva, Phys. Rev. Lett. **92**, 027201 (2004).

7 W. Stoecklein, S. S. P. Parkin, and J. C. Scott, Phys. Rev. B **38**, 6847 (1988).

8 J. McCord, R. Mattheis, and D. Elefant, Phys. Rev. B **70**, 094420 (2004).

9 B. K. Kuanr, R. E. Camley, and Z. Celinski, J. Appl. Phys. 93, 7723 (2003).

10 P. J. H. Bloemen, H. W. van Kesteren, H. J. M. Swagten, and W. J. M. de Jonge, Phys. Rev. B **50**, 13505 (1994).

11 K. J. Lee, O. Redon, and B. Dieny, Appl. Phys. Lett. **86**, 022505 (2005).

12 S. Mangin, D. Ravelosona, J. A. Katine, M. J. Carey, B. D. Terris, and E. E. Fullerton, Nature Materials **5**, 210 (2006).

13 S. S. Parkin, N. More, K. P. Roche, Phys. Rev. Lett. 64, 2304 (1990).

14 Chantrell R W, Wongsam M, Schrefl T and Fidler J 2001 *Encyclopedia of Materials: Science and Technology* ed Buschow K H J, Cahn R W, Flemings M C, Ilschner B, Kramer E J Mahajan S (Elsevier) 5642-5651

15 Schrefl T, Fidler J, Chantrell R W and M. Wongsam, 2001 *Encyclopedia of Materials: Science and Technology* ed Buschow K H J, Cahn R W, Flemings M C, Ilschner B, Kramer E J. Mahajan S (Elsevier) 5651-5661

16 J.C. Slonczewski, J. Magn. Magn. Mat. **247**, 324-338 (2002).

17 G. Bertotti, C. Serpico, I.D. Mayergoyz, R. Bonin, M. d´Aquino, J. Magn. Magn. Mat. **316,** 285-290 (2007)

18 W. Scholz, T. Schrefl, and J. Fidler, JMMM 233, 296-304 (2001)

19 W.F. Brown, Jr., Micromagnetics (Wiley, New york, 1963)

20 V.D.Tsiantos, T. Schrefl, W. Scholz, and J. Fidler, J. Appl. Phys. 93, 8576-8578 (2003)

21 F.J.A. den Broeder, E. Janssen, W. Hoving and W.B. Zeper, IEEE Trans. Magn. 28, 2760 (1992).

22 Joo-Von Kim, Phys. Rev. B 73, 174412 (2006).

List of figures:

Figure 1. (Color online) Calculated output voltage of a Co(30nm)/Ru(0.8nm)/Co(3nm) nanopillar for a current of 36 mA. The width and length are 80 nm and 180 nm, respectively. The output voltage was calculated for a temperature of 0 K and 450 K

Figure 2. (Color online)Calculated power spectrum density of a Co(30nm)/Ru(0.8nm)/Co(3nm) nanopillar. The width and length are 80 nm and 180 nm, respectively

Figure 3. (color Online) Oscillator frequencies as function of current of a Co (30nm) / Ru (0.8nm) / Co(3nm) nanopillar at 0 and 450 K.

Figure 4. (Color online) Power spectrum density as function of the current for the oscillator with perpendicular free layer at T = 0 K.

Figure 5. (Color online) Power spectrum density as function of current (3 to 5 mA) and temperature (0 to 450 K) for an oscillator with perpendicular magnetized active (free) layer.

Figure 6. (Color online) Left hand side: Line width as function of current. Right hand side: Line width as function of spectral line density (right) for an oscillator with perpendicularly magnetized active (free) layer.

Figure 7. (Color online) Voltage output as function of time. A slight shift to higher frequencies occurs with increasing temperatures.

Figure 8. (Color Online) Left hand side: Frequency as a function of the current for a circular nano-pillar . There are no external field sources. Right hand side: Full width half maximum as a function of the current.

Figure 9. (Color online) Power spectrum density for a current of 12 mA and 13 mA at temperatures of 300 K and 450 K.

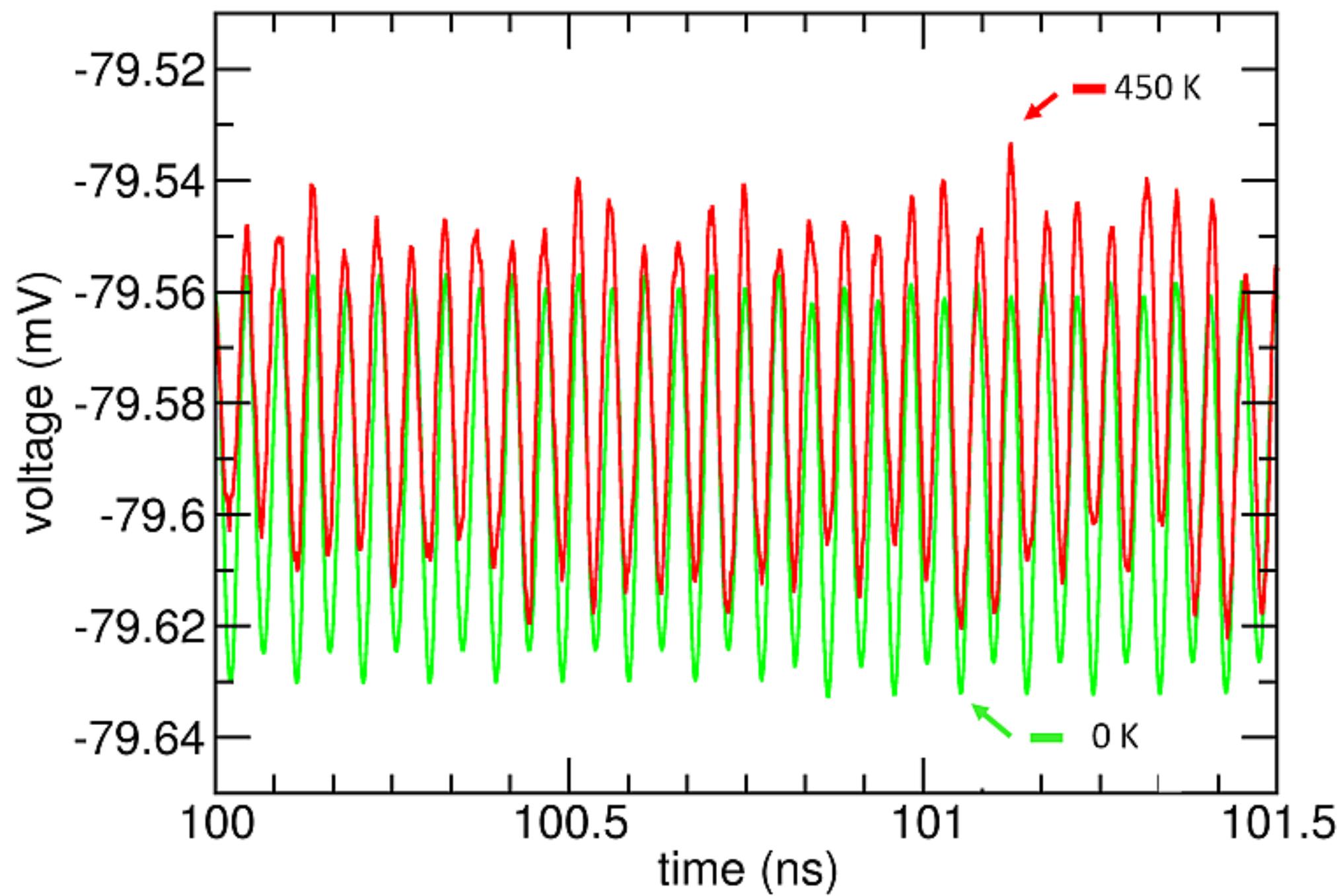

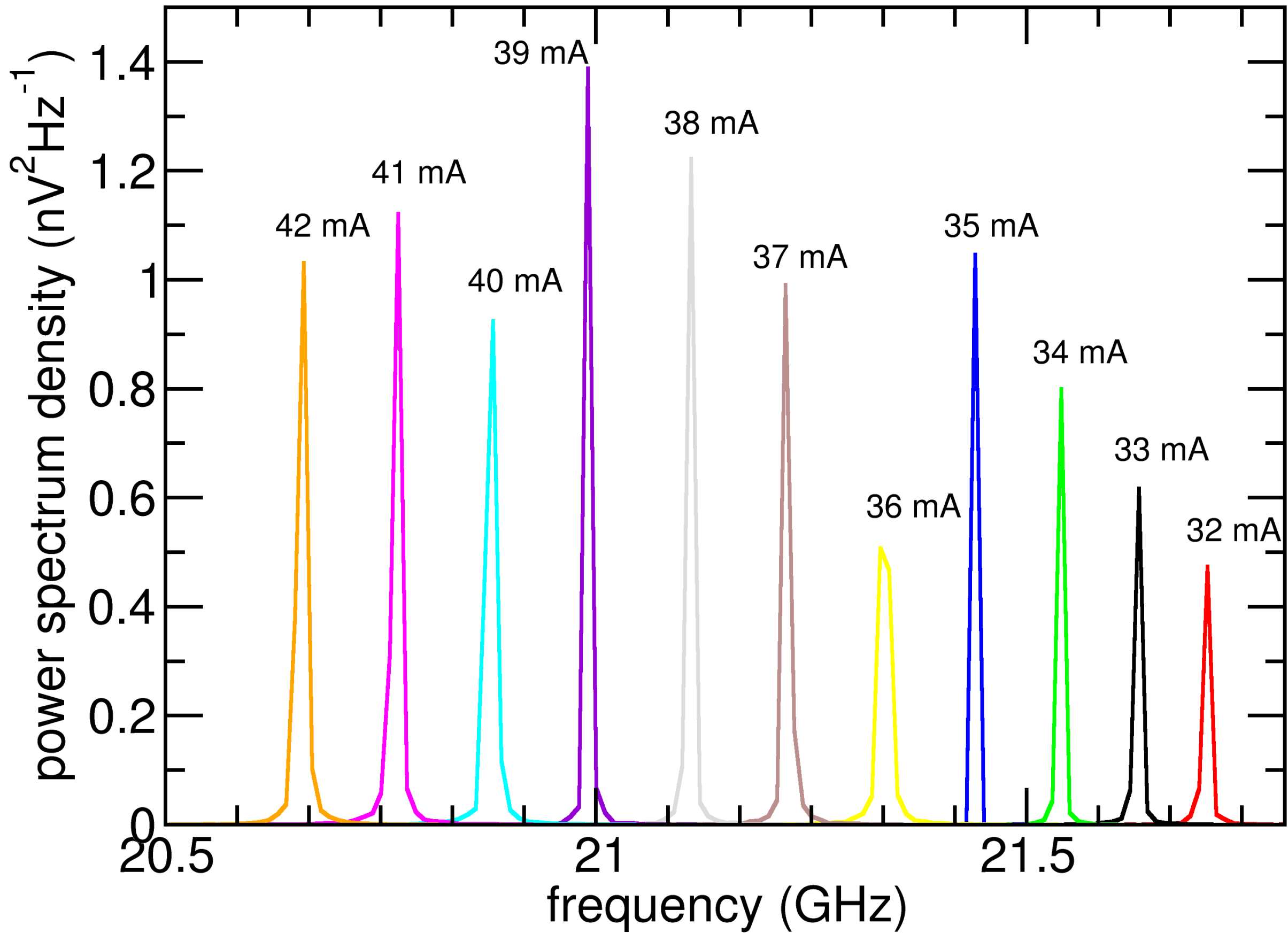

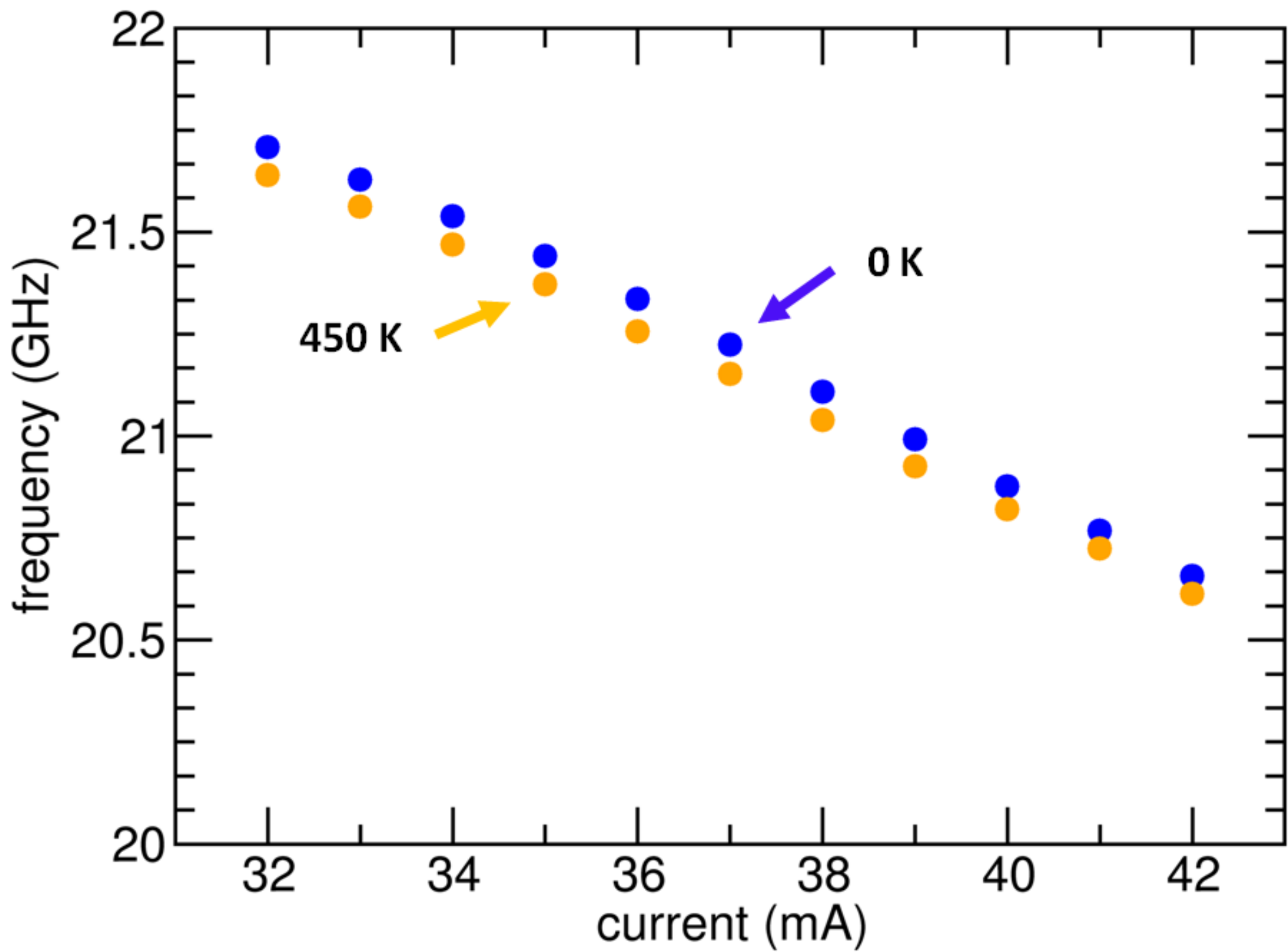

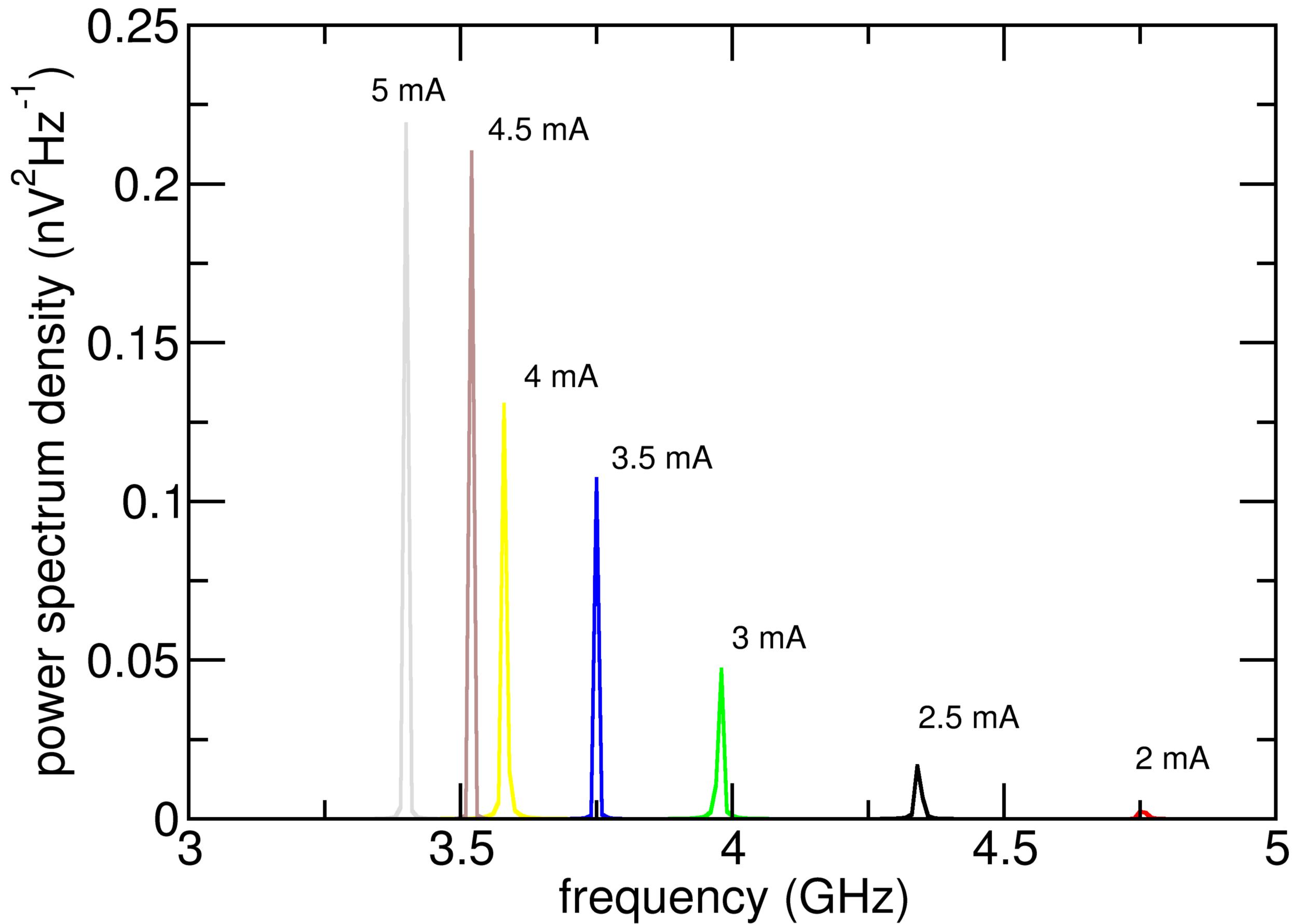

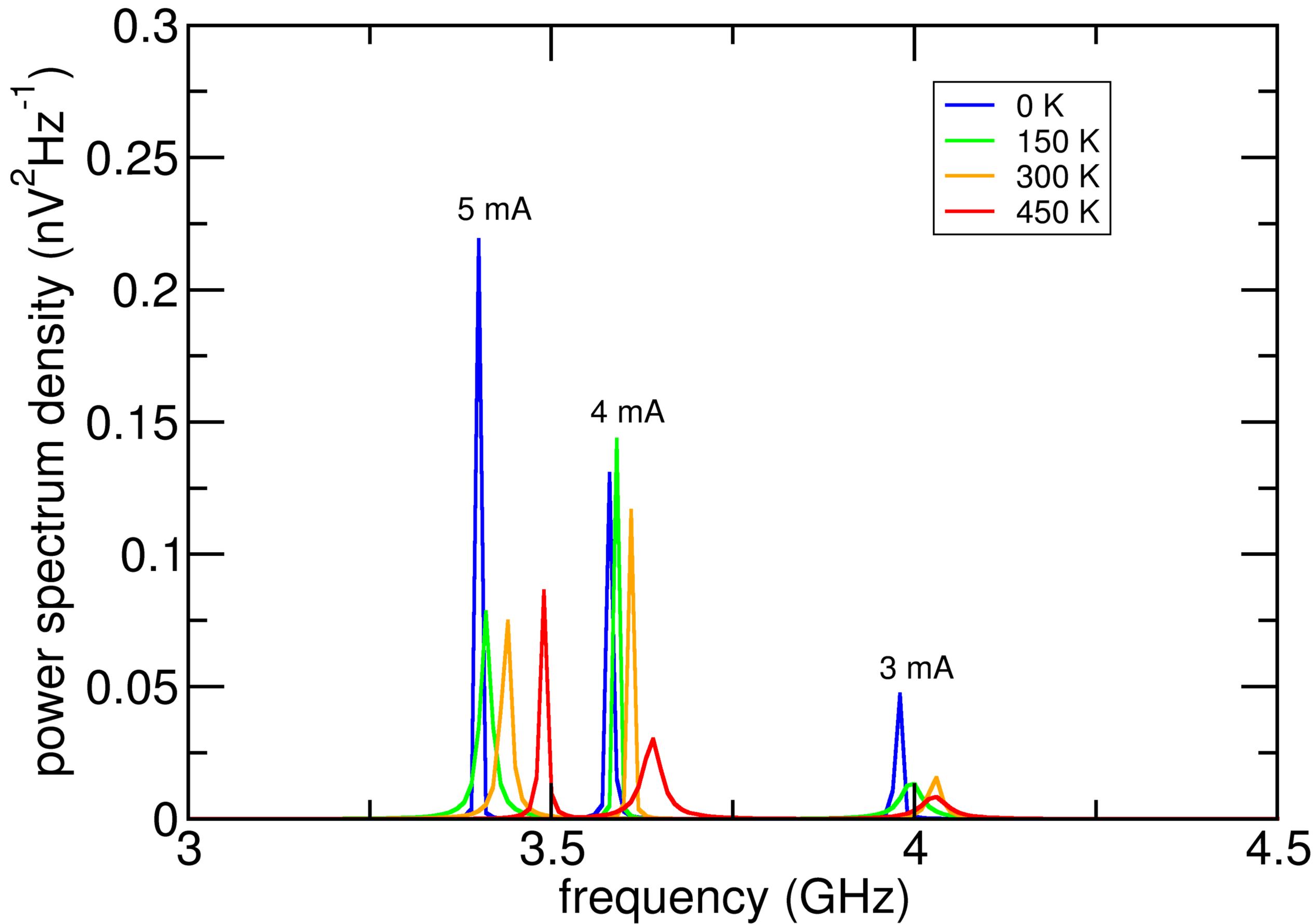

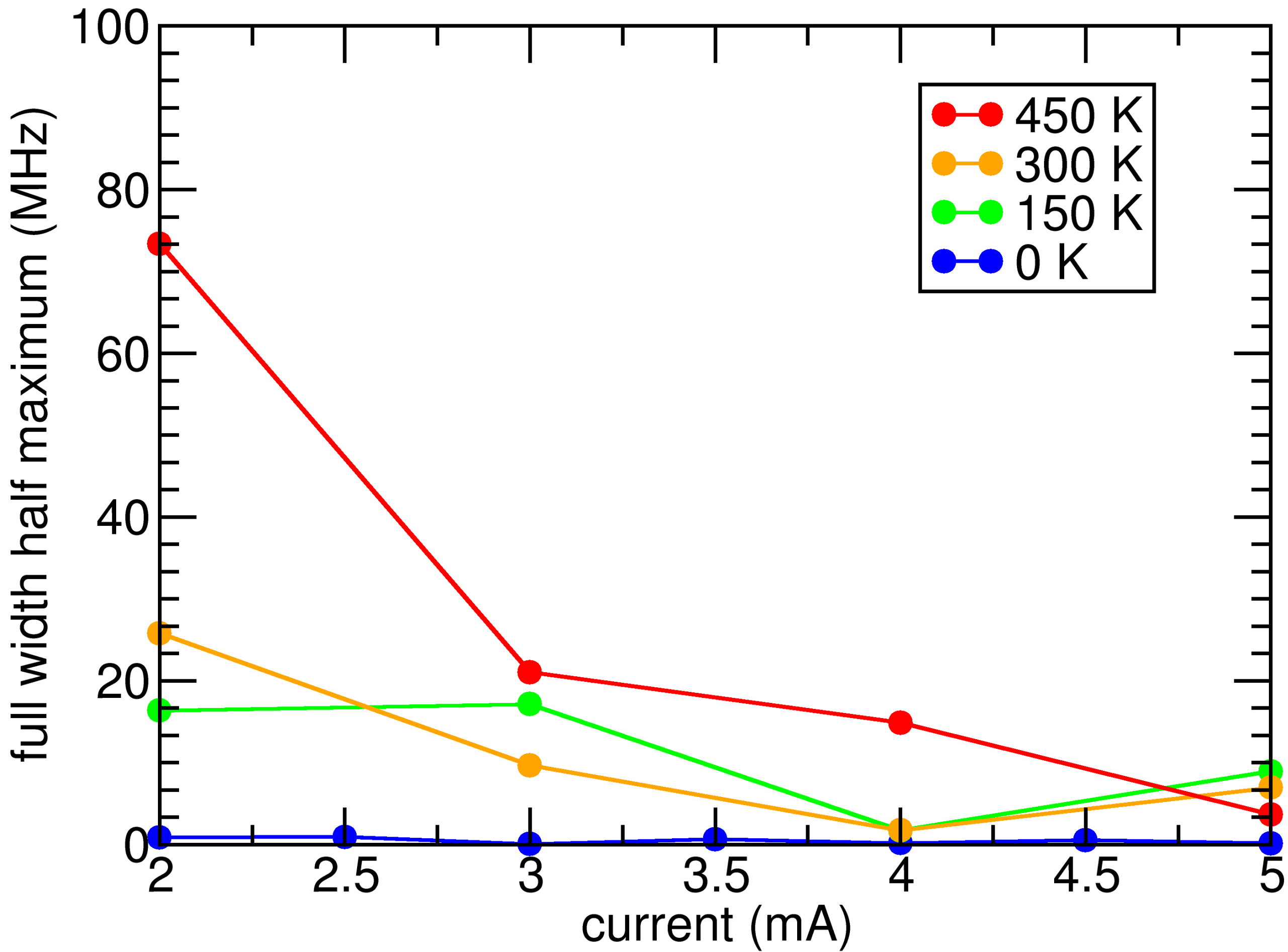

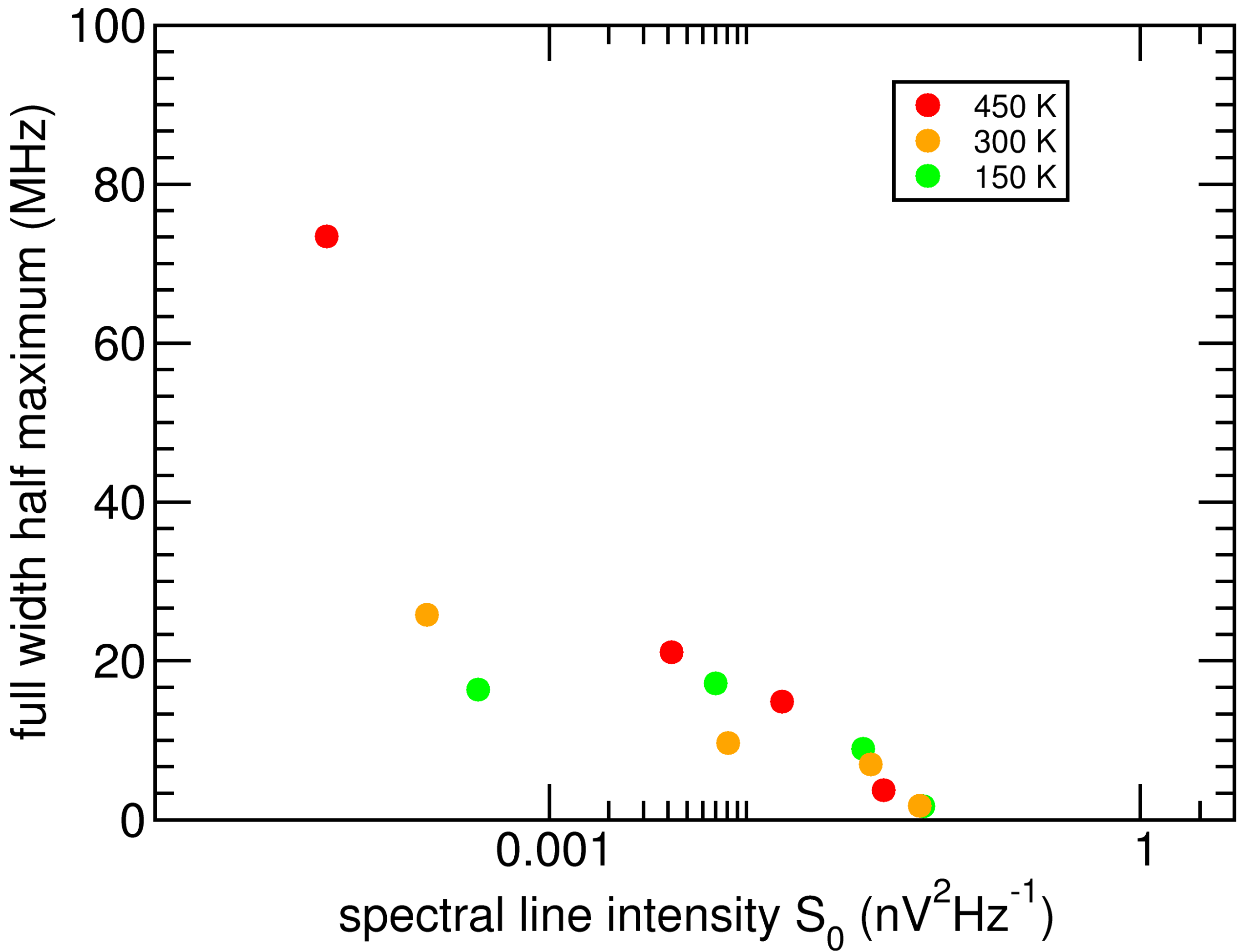

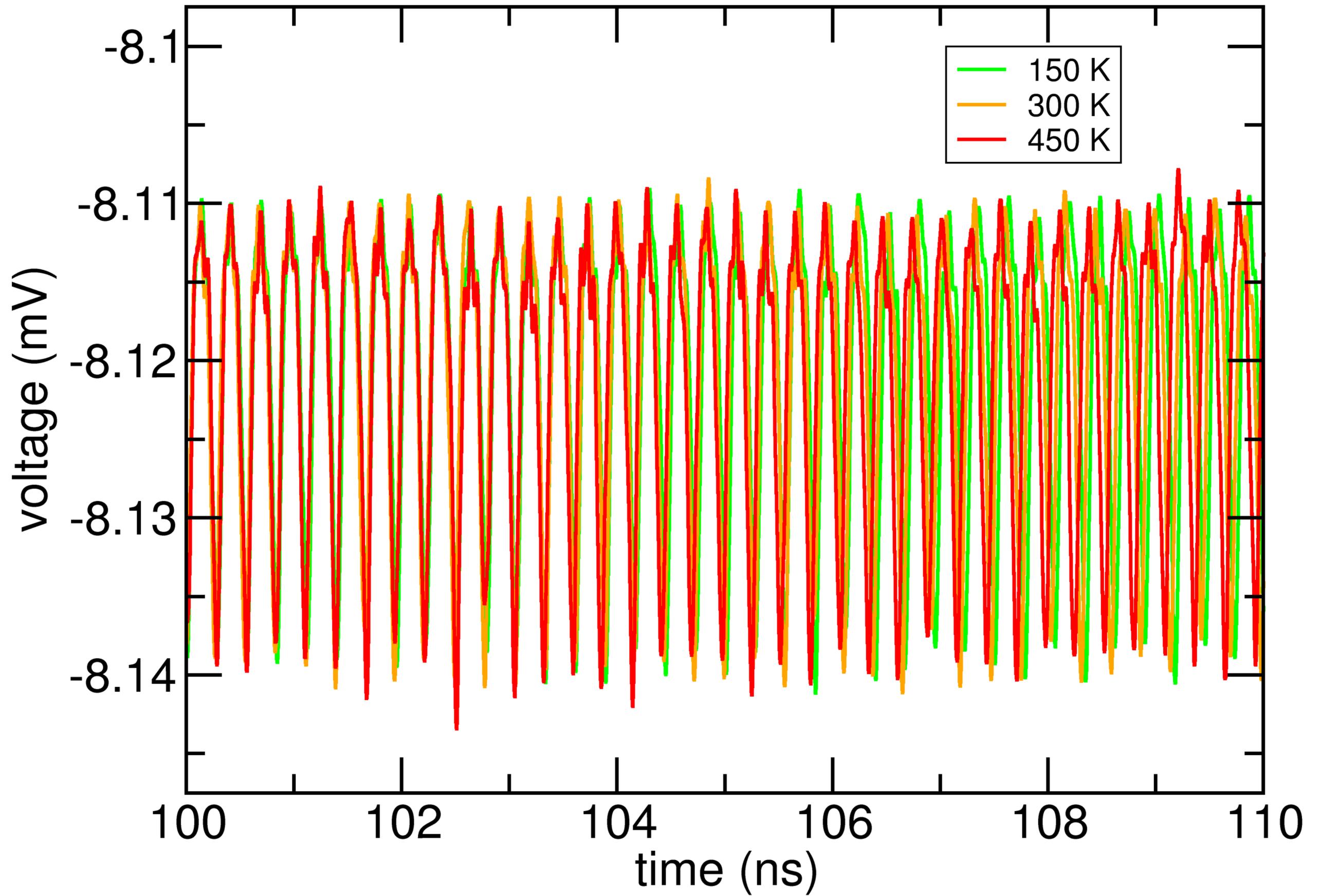

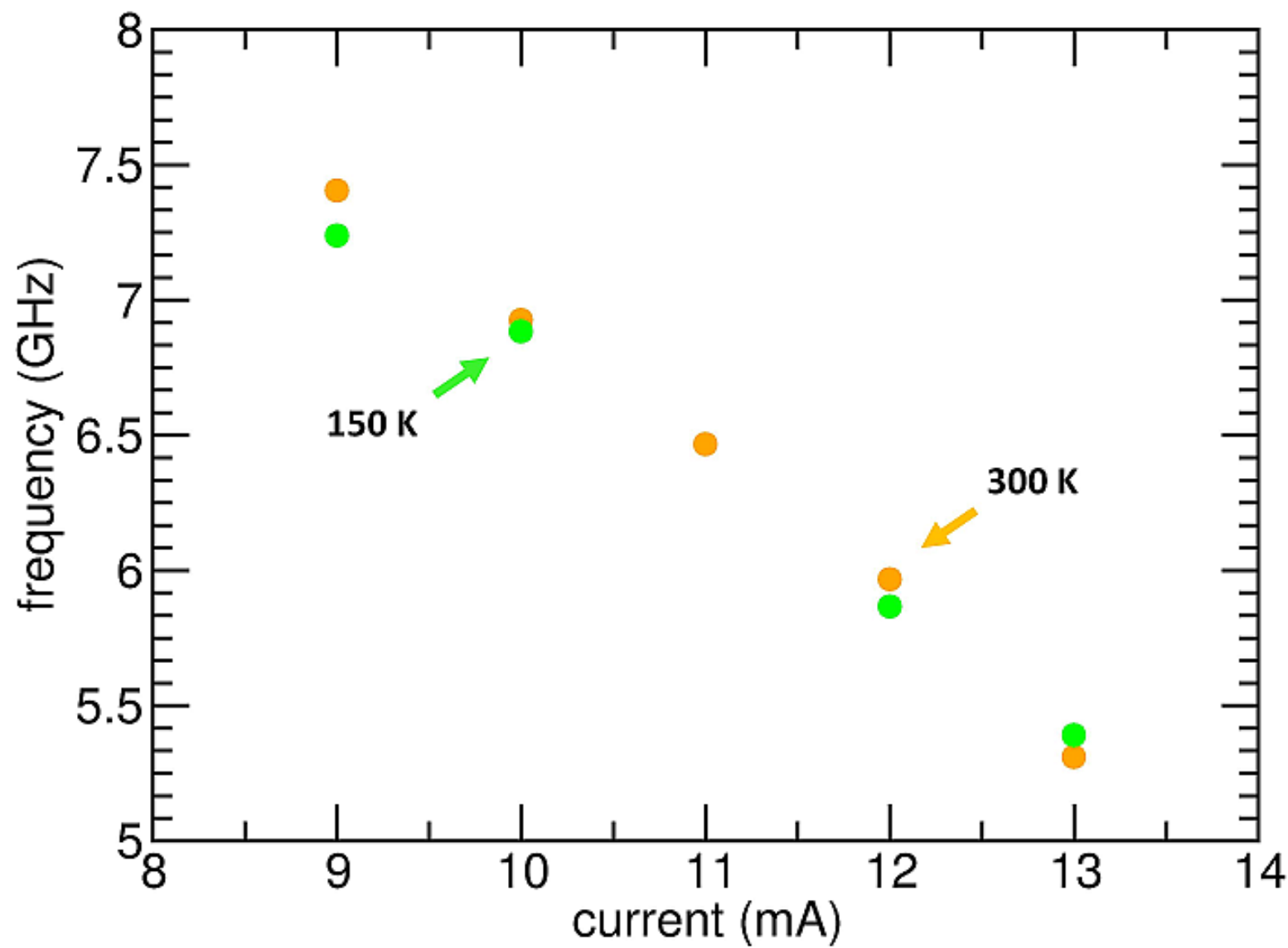

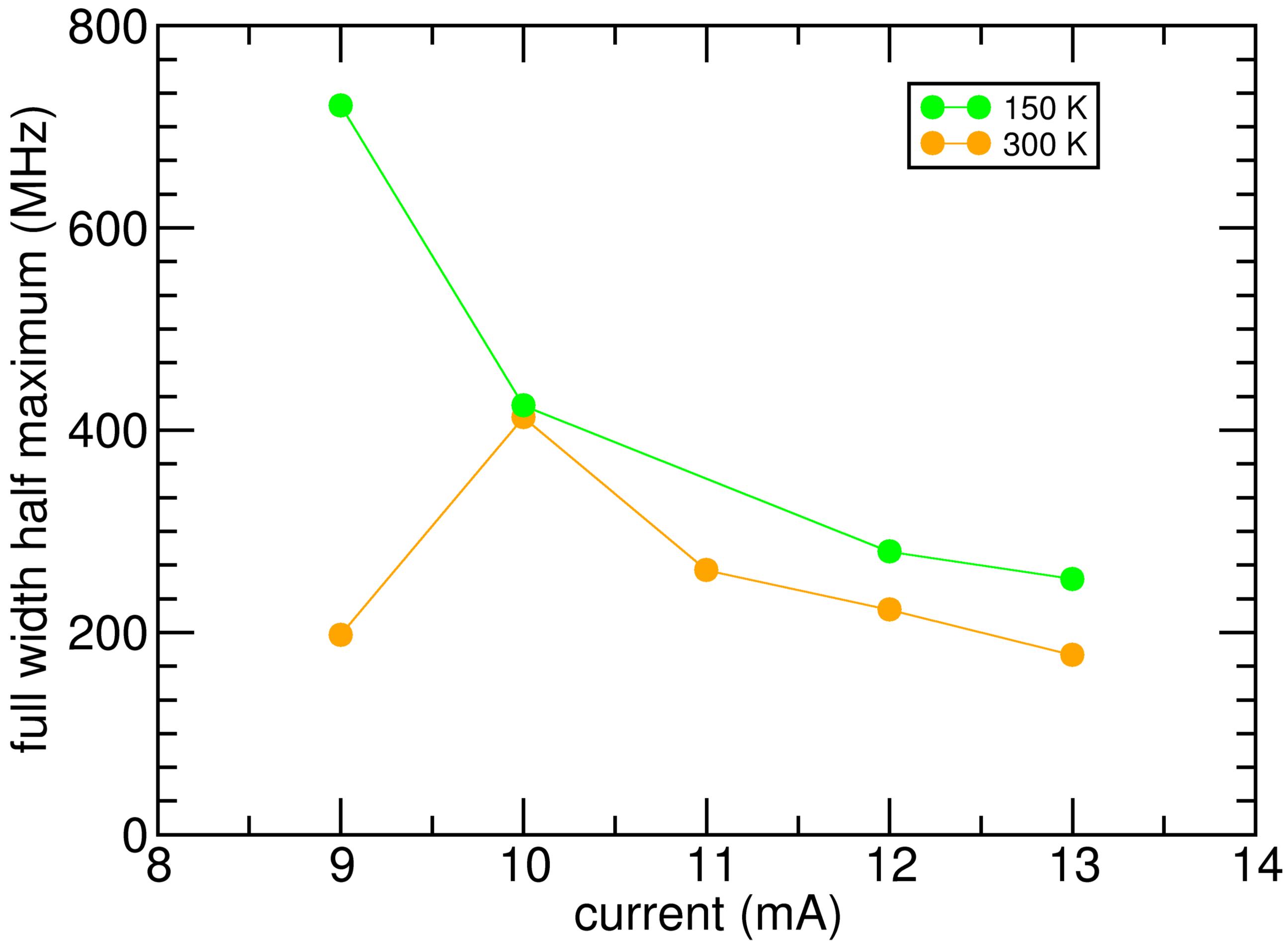

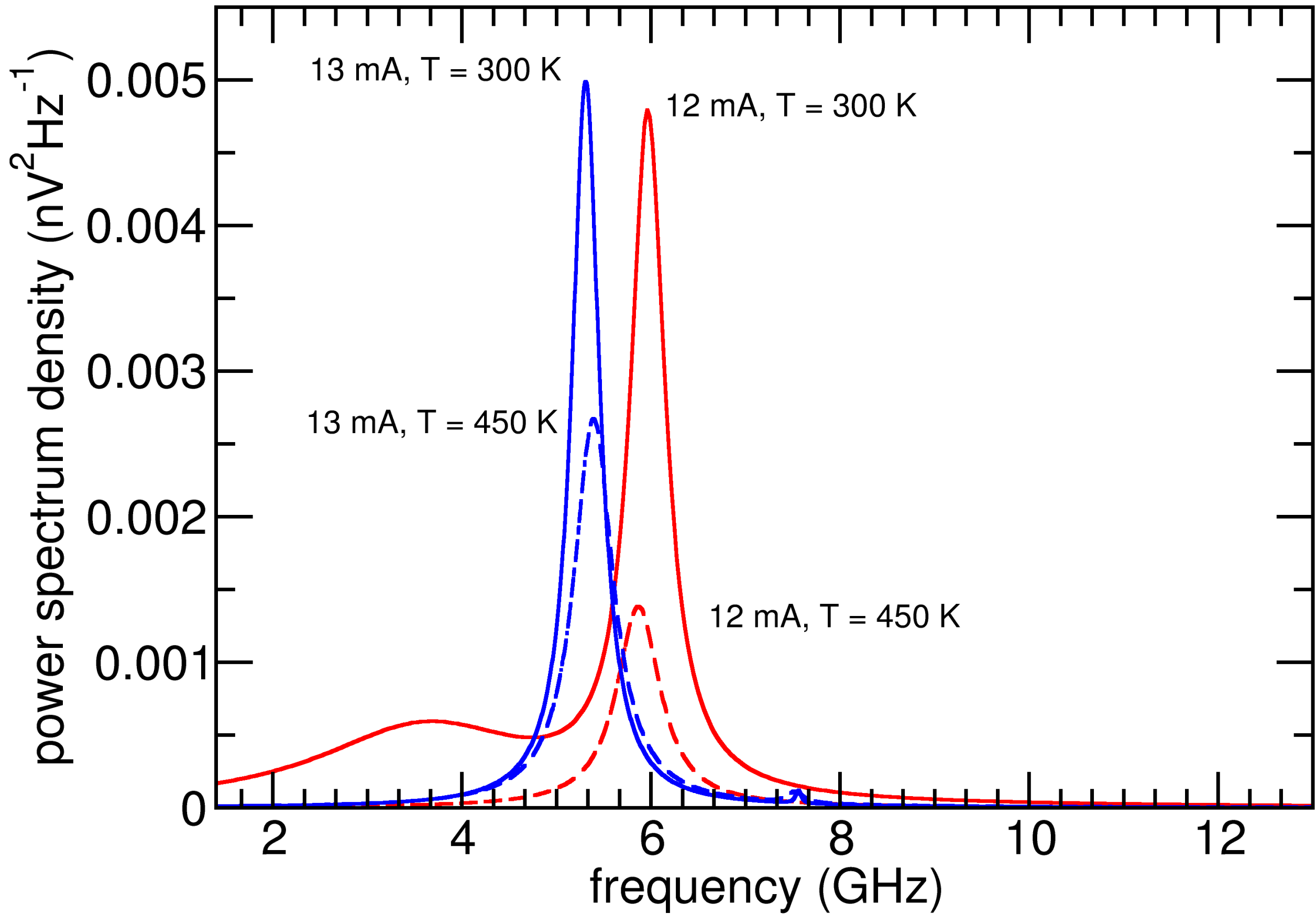